\documentclass[universe,article,accept,pdftex,moreauthors]{mdpi} 

\firstpage{1} 
\pubvolume{1}
\issuenum{1}
\articlenumber{0}
\pubyear{2024}
\copyrightyear{2024}
\externaleditor{Academic Editor: Firstname Lastname}
\datereceived{ } 
\daterevised{ } 
\dateaccepted{ } 
\datepublished{ } 
\hreflink{https://doi.org/} 

\Title{The effect of outflow launching radial efficiency of accretion disk on the shape of emission-line profiles}

\TitleCitation{The effect of outflow on the BLR}

\Author{
Mohammad Hassan Naddaf $^{1}$\orcidA{}}

\AuthorNames{Mohammad Hassan Naddaf}

\AuthorCitation{Naddaf, M. H.}

\address{%
$^{1}$ \quad Institut d'Astrophysique et de Géophysique, Université de Liège Allée du six août 19c, B-4000 Liège (Sart-Tilman), Belgium}

\corres{Correspondence: mh.naddaf@uliege.be}

\abstract{This paper presents a preliminary investigation into the influence of radial behavior of disk outflow on the structure and dynamics of the broad line region (BLR) in active galactic nuclei (AGNs), with the emphasis on how the mass ejection rate contribute in shaping the broad emission line profiles. Specifically, we analyze how varying the radial efficiency of mass loss from accretion disks, driven by radiative dust-based mechanisms, contribute to the distribution of material in the BLR. By exploring different radial scenarios of disk mass loss behavior, we uncover connections between outflow radial efficiency and emission line profiles, particularly for lowly ionized lines. Our findings reveal that while the observed shape of broad emission lines is partially influenced by the radial behavior of the disk outflow, it ultimately depends more critically on the physical conditions of the clouds and the specific approach adopted to the emissivity for their contribution to the line formation.}


\begin{document}


\section{Introduction}

AGNs are among the most luminous and energetic objects in the universe, powered by accretion of matter onto supermassive black holes at their centers. AGNs are characterized by their emission across the electromagnetic spectrum, and one of the most critical diagnostic features of AGNs is the presence of emission lines in their spectra \citep{schmidt1963, antonucci1993, netzer2015, Almeida2017}. These emission lines provide vital clues to the physical conditions and kinematics of the emitting region. The emission lines in AGN spectra are typically divided into two categories: narrow emission lines and broad emission lines, which originate from different regions of the AGN environment. The BLR which is closer to the central black hole, gives rise to the broad emission lines observed in the optical and ultraviolet spectra. These lines are broadened due to the high velocity of gas in the BLR, which moves at speeds of thousands of kilometers per second due to the intense gravitational influence of the black hole \citep{gaskell2009}. The broad emission lines in AGNs exhibit complex line profiles that provide critical information about the dynamics and structure of the BLR. The BLR is thought to be stratified, with different ionization states present at different distances from the black hole. This stratification gives rise to the observed differences in the profiles of high-ionization lines (e.g., C IV, and He II) and low-ionization lines (e.g., H$\beta$, and Mg II) \citep{CollinSouffrin1988}. High-ionization lines originate closer to the central source where the gas is more highly ionized, while low-ionization lines are produced further out \citep{kollatschny2003, czerny2015}. The line profiles of these emission lines reflect the dynamics of the BLR, with Doppler broadening due to the high velocity of the gas, and asymmetries in the profiles providing evidence for outflows, inflows, or other kinematic complexities \citep{leighly2007}. The BLR is a key region in the AGN environment, and understanding its structure is crucial for interpreting the observed emission lines. The gas in the BLR is believed to be in Keplerian orbits around the supermassive black hole, with ionization stratification across the region \citep{collin2006}. Studies suggest that the BLR is not a homogeneous entity but consists of clouds or filaments of gas in various ionization states \citep{kollatschny2003}. Recent models propose that accretion disk winds and radiation pressure may play a significant role in shaping the kinematics of the BLR. 
Recent advancements in high-resolution spectroscopy and interferometry have provided unprecedented insights into the radial dynamics of gas in AGNs, revealing complex structures within the BLR, including variable disk outflows \citep{GravityColl2018, honig2013}. Among the key processes proposed to drive outflows in the BLR is radiative dust-driven mechanisms, especially in lowly ionized regions where dust can survive at large distances from the AGN. These outflows may carry a significant portion of mass from the accretion disk, affecting the radial density and ionization profiles within the BLR and, consequently, the observed line profiles \citep{elitzur2006}. The concept of disk winds in AGNs, initially developed within the framework of broad absorption lines, has been extended to BLRs, suggesting that mass loss rates and wind geometry may also play a pivotal role in shaping emission line profiles.
Studies have shown that the structure of the BLR is highly sensitive to the radial behavior of disk outflows \citep{proga2000, naddaf2021}. The radial function of outflow launching efficiency is supposed to be a key factor determining whether gas remains optically thick or thin, influencing emission line shapes through its impact on the kinematics and density structure of the BLR clouds \citep{netzer1993, nicastro2000, gaskell2009}. Such influences offer a compelling explanation for the variability in line profile asymmetries and peak shifts observed in low-ionization lines such as H$\alpha$, H$\beta$, and Mg II, which are believed to originate from denser and more dust-enriched zones within the BLR \citep{czerny2011, naddaf2022}.

Despite extensive theoretical models of BLR structure, there is ongoing debate on the how mechanisms launching outflow influencing the mass loss from the disk and density distribution of material in the region. A critical issue is the dependency of BLR geometry and line profile shapes on radial mass loss functions, which determine the density stratification above the disk \citep{murray1995}.
In this paper, we intend to investigate the effect of disk outflow behavior on the shape of emission lines by systematically examining different radial mass loss scenarios and analyzing their implications for the distribution of material and geometry of the BLR in AGNs. We focus on lowly-ionized emission lines, which are particularly sensitive to dust-driven forces, and propose various scenarios that reflect physically-motivated disk outflow behaviors. In order to do so, we use our model for the formation of low-ionization BLR which is based on the radiatively dust-driven mechanism; more specifically, known as FRADO (Failed Radiatively Accelerated Dusty Outflow) model. The paper is structured as follows. We provide a concise overview of the FRADO model and its mechanism for the formation of the BLR in section \ref{sec:FRADOmodel}. This is followed by section \ref{sec:massloss} which details the approaches to the disk outflow behavior. Next, we present the results of emission line profiles under two different outflow scenarios in section \ref{sec:results}. Finally, the results are discussed in section \ref{sec:discussion}, and conclusions are drawn at the end.

\section{FRADO model}\label{sec:FRADOmodel}

The FRADO  model, developed over the last decade \citep{czerny2011, czerny2015, czerny2016, czerny2017}, presents a unique approach to understanding the interaction between radiation pressure and dusty material lifted from the surface of an accretion disk. Unlike conventional wind models, FRADO focuses on a "failed wind" scenario, where dusty clumps are ejected from the disk surface in a vertical direction but do not escape to infinity. This simplified one-dimensional framework excludes the orbital dynamics of the material, narrowing its scope to vertical motion alone. At the heart of this model is the idea that the surface layers of the cold accretion disk, rich in both dust and gas \citep{rees1969, dong2008}, are exposed to local radiation. This radiation flux is powerful enough to lift dusty clumps away from the disk. However, as these clumps rise to greater heights, they encounter stronger radiation from the central black hole's vicinity, which strips them of their dust. The remaining gas clumps, now devoid of dust, fall back towards the disk in a ballistic trajectory, governed by the gravitational pull of the black hole.

The FRADO model is one of the few frameworks that physically-motivated self-consistently integrates the role of dusty winds in BLR formation, without introducing any arbitrary free parameters. Its strength lies in its ability to explain the inner radius of the BLR as the dust sublimation radius, which has been supported by reverberation mapping studies \citep{naddaf2020}. Additionally, the model provides a physical mechanism for outflows in AGN without relying on magnetohydrodynamic effects, instead focusing on radiative driving. The inner radius, or actually the onset of BLR, and the span of radii capable of launching (failed) wind is not arbitrarily but a self-consistently output of the model \citep{czerny2017}.
The model has been then further refined in recent years, upgraded to 2.5D version, offering detailed insights into how radiation pressure can drive the formation and motion of BLR clouds. The new version folds the proper prescription of the disk radiation pressure with wavelength dependent dust opacities and also the role of shielding effect to provide a plausible picture of low-ionization BLR. However, possible limitations can be present due to, for example, dust grain properties which are not well constrained observationally, the focus of the model due to its nature on the low-ionization part of BLR (thereby no prediction for the highly-ionized lines such as CIV or such), or the underlying standard Shakura-Sunyaev (SS) disk model \citep{SS73} which makes it limited to Eddington ratios between 0.01 up to near Eddington. However, the new version of the model has been very successful in explaining many aspects of observation of AGNs including for example the overall shape of low-ionization line profiles, formation of escaping outflow, and the spread in the radius-luminosity relation \citep[see,][for more details]{naddaf2020, naddaf2021, naddaf2022, naddaf2022Dynam}.

FRADO model is essentially based on the radiatively efficient thermal (blackbody) emission in the optical/UV bands from the standard thin accretion disk \citep{SS73} operating at higher Eddington ratios [0.01 to 1] than that of radio-loud (jetted) AGNs (around or below 0.01) where accretion flow may transition to advection-dominated accretion flow (ADAF) or magnetized thick disks. In jetted AGNs, the jets dominate the feedback processes and disk winds are often suppressed due to the competition between jets and wind-launching mechanisms; while in radio-quiet ones radiatively driven winds are more prominent, leading to broader absorption/emission lines and observable outflows in the UV and X-rays. Overally the FRADO model as namely implied, belongs to AGN systems where the radiative output is dominated by thermal processes rather than non-thermal synchrotron radiation, so that SS disk is present and radiative dust-driving wind launching is efficient. Therefore in the context of FRADO by definition, non-jetted luminous AGNs including Seyferts and radio-quiet quasars with SS underlying disk are the case.

\section{Formulations of the disk outflow rate}
\label{sec:massloss}

The radial behavior of the disk outflow\footnote{In this paper, the phrases outflow, wind, and mass loss are used interchangeably.}rate $\dot{M}_z(r)$ (where the subscript $z$ is just to remind the quasi-vertical nature of the outflow from the disk), defined as the mass loss rate per unit radius due to outflows, is expected to follow a power-law profile of the form $\dot{M}_z(r) \propto r^{s}$. In this context, the power-law index $s$ is anticipated to be negative, reflecting the physical reality that the capability of the disk to launch outflows should diminish with increasing disk radius. This behavior aligns with theoretical models and observational evidence suggesting that radiative forces, which are key drivers of outflows, are most effective in the inner regions of the disk. Such scaling laws have been explored in various contexts, including disk winds in AGNs \citep{MurrayChiang1995}, dust-driving winds \citep{naddaf2022}, and line-driven outflows in stellar atmospheres \citep{castor1975}. The ejected material forming a wind, either failed or fully escaping outflow, is supposed to be responsible for the formation of the BLR \citep[e.g.,][]{czerny2011, naddaf2021}. A negative $s$ reflects a physically plausible scenario, as it captures the rapid drop in flux, $F(r) \propto r^{-3}$ (or, $L(r) \propto r^{-2}$, in terms of the local luminosity), with increasing radius in geometrically thin, optically thick accretion disks \citep{SS73}. 

To characterize the local mass loss rate from the surface of an accretion disk, in the dust-dominated region of interest of this paper where the dust-driving mechanism is favored more, we start with the single-scattering. Thereby, the clumps formed at the disk surface should be approximated as optically thin \citep{czerny2017}, and the momentum of radiation is transferred to the outflow, as in the theory of stellar winds \citep[see][and the references therein]{muijres2011}. Therefore, following \citep{arav1994}, we let the energy flux $F(r)$ at a radius $r$ from the disk to be proportional to the vertical mass outflow rate $\dot{m}_z$ as below

\begin{equation}
\frac{F(r)}{c} \propto \dot{m}_z \, v_{\text{f}}
\end{equation}
where $\dot{m}_z $ is the vertical mass loss rate from the disk per unit area, \( v_{\text{f}} \) is the terminal velocity for material at the disk surface at radius \( r \) which we take it to be equal to $\sqrt{GM/r}$, i.e. the local escape velocity \citep{lamers1995}, $ c $ is the speed of light, and \( F(r)= \frac{3 G M \dot{M}}{8 \pi r^3}\) is the radiative flux from the disk per unit area at radius $r$ (as the dust-driving mechanism works at large radii, we do not include the inner boundary condition) where $\dot{M}$ is the intrinsic accretion rate of the disk which is constant, $G$ is the gravitational constant, and $M$ is the black hole mass. The component of gravity at large radii at the disk surface is negligible ($H/r << 1$). So the local mass loss rate per unit area of disk surface is

\begin{equation}
\dot{m}_z \propto r^{-5/2}
\end{equation}

This, if all other constant and parameters are carefully addressed, can be integrated over a specified radial range to give the cumulative radial mass loss rate (in optically thin regime approximation), however, part of the ejected mass eventually returns to the disk in the form of failed winds \citep{czerny2011, naddaf2021}. But the local mass loss rate\footnote{$\dot{M}_z(r) = d\dot{M}_z/dr$, where $d\dot{M}_z = \dot{m}_z \cdot 2\pi r dr$} from the disk surface as a function of radius $r$, denoted by $\dot{M}_z(r)$, can be given as

\begin{equation}\label{eq:massloss}
\dot{M}_z(r) \propto r^{-1.5}
\end{equation}

The equation \ref{eq:massloss} shows how outflow launching efficiency scales radially across the disk, with larger mass loss expected in regions where the flux is higher or the escape velocity is lower. But the decay of the outflow intensity with radius is shallower than what reported for quasars \citep{murray1995, MurrayChiang1995}, where outflow intensity follows a steeper radial decline.

On the other hand, in the optically thick regime, where multiple scatterings are allowed, we preferably follow an energy-wise approach, rather than the momentum-oriented solution above, as the entire energy can be transferred to the outflow. Through this approach, we also incorporate the role of dust opacity in the lowly-ionized regions of the BLR, where radiative dust-driving is a significant mechanism. In regions where dust is present, the radiative flux driving the outflow interacts with dust grains more efficiently than electrons, so
\begin{equation}
F(r) \frac{\kappa_{d}(r)}{\kappa_{T}} \propto \dot{m}_z \, U_{\text{g}}
\end{equation}
where $U_{\text{g}}$ is the gravitational potential per unit mass, $\kappa_{T}$ is the Thompson scattering opacity, and $\kappa_{d}(r)$ is the dust opacity which is expected to scale with radius as $r^{\alpha}$. 
This modification yields
\begin{equation}
\dot{M}_z(r) \propto r^{\alpha-1}
\end{equation}
where $\alpha$ is negative $\sim -0.5$ \cite[see e.g.,][]{Semenov2003, naddaf2021}.
Now in order to investigate how these different decay rates influence the geometry and distribution of material in BLR, consequently, the resulting shapes of broad emission lines, we proceed with adopting outflow rate power-law indices of $s=-0.5$ and $-1.5$.

\section{Results}\label{sec:results}

To investigate the effects of the radial scaling of disk outflow rates on the geometry and distribution of BLR clouds, we employed 2.5D FRADO computations. In order to do so, the model only requires the physical parameters of the source, including black hole mass, accretion rate, dust sublimation temperature, dust model of interest, but no any free arbitrary parameter. The inner/outer launching radii for the simulations in 2.5D FRADO is not an input, but is self-consistently found as an output based on the dust-sublimation temperature and the radiative dust-driving outflow launching efficiency for a given radius \citep[see][for the details]{naddaf2021}.

As a benchmark for comparison, we adopt the physical parameters of the mean composite quasar \citep{vandenberk2001} from the Sloan Digital Sky Survey (SDSS), for which the black hole mass is $\log(M_{\rm BH}/M_{\odot}) = 9.1$, and the accretion rate is $\log(\dot{M}/ \dot{M}_{\rm Edd}) = -0.9$. It can represent quasars in SDSS data well \citep{shen2011}. The benchmark is defined by the single-peaked shape of the low-ionization broad emission lines observed in the mean composite quasar spectrum \citep{vandenberk2001}, providing a baseline for comparison with the line profiles derived from the different radial behavior of disk outflow rates.

\begin{figure}[H]
\begin{adjustwidth}{-\extralength}{0cm}
\centering
\includegraphics[width=16cm]{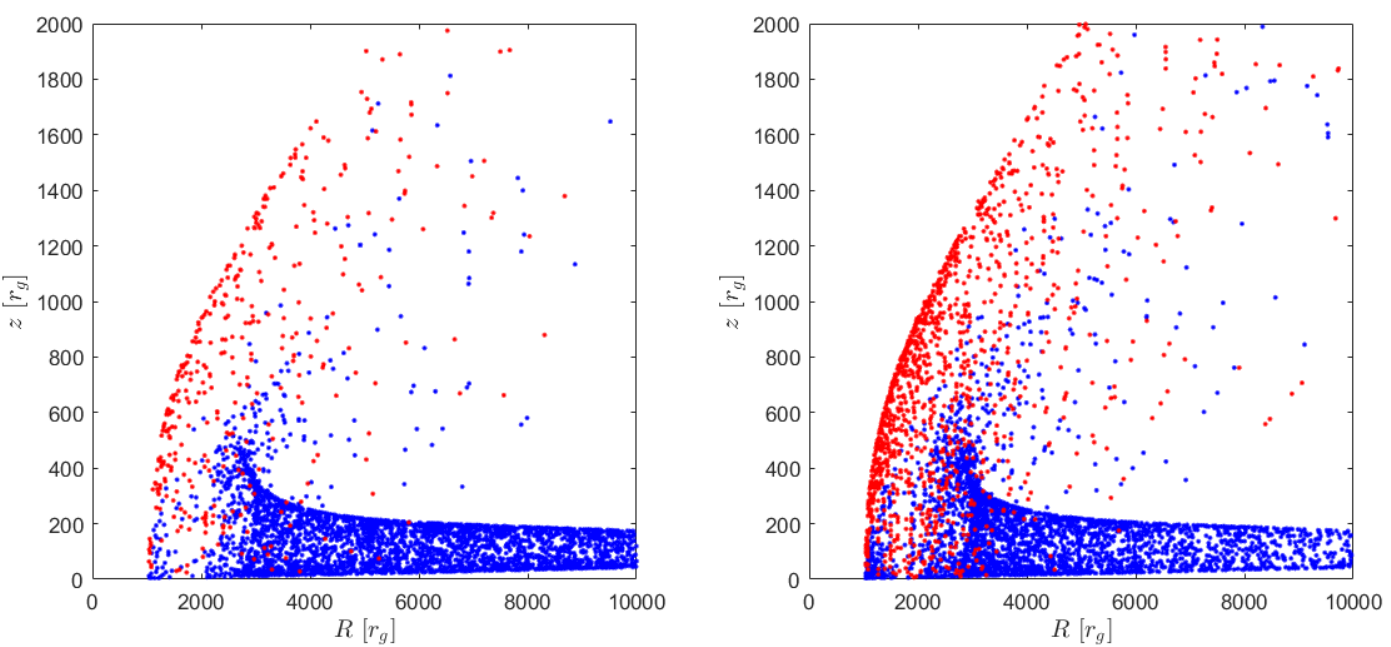}
\end{adjustwidth}
\caption{Projected perspective of the spatial distribution of BLR clouds for the mean quasar model, based on the two adopted radial outflow rate functions. The left panel represents the shallower outflow scaling with $s=-0.5$, while the right panel illustrates the steeper scaling with $s= -1.5$. The horizontal and vertical axes denote the radial position and height of the clouds, expressed in units of the gravitational radius, $r_{\rm g}$, of the mean quasar's black hole. Dusty clouds are shown in blue, and dust-free clouds are displayed in red.\label{fig:distribution}}
\end{figure}

The spatial distribution of BLR clouds resulting from two different radial functions of the outflow launching efficiency is shown in figure \ref{fig:distribution}. For each model, the locations and velocities of the BLR clouds are computed in 3D Cartesian coordinates using the 2.5D FRADO. These distributions highlight the role of outflow efficiency radial scaling in shaping the BLR geometry and dynamics. Specifically, the shallower scaling gives a more extended BLR with slower-moving clouds distributed over a broader range of radii, while the steeper one produces a more compact BLR with clouds concentrated closer in to the black hole and exhibiting higher velocities.

\begin{figure}[H]
\begin{adjustwidth}{-\extralength}{0cm}
\centering
\includegraphics[width=17cm]{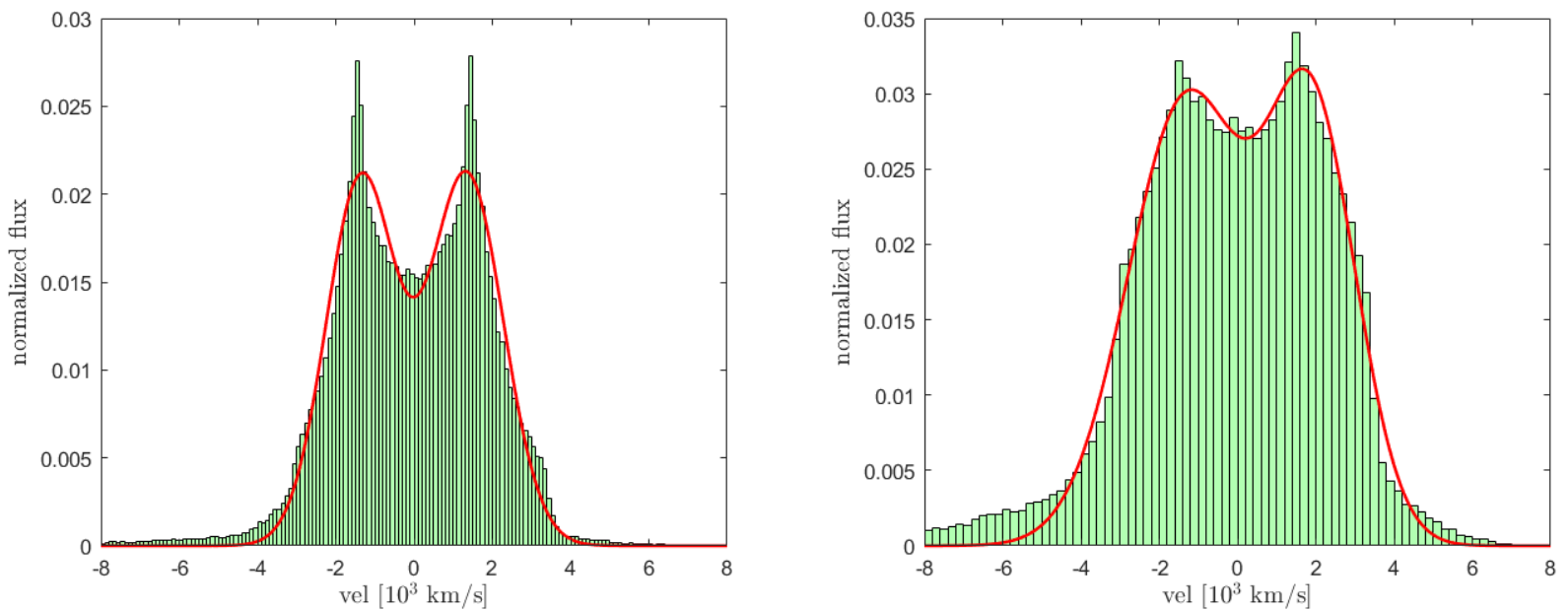}
\end{adjustwidth}
\caption{The emission line profiles for the model of mean quasar observed at the mean viewing angle of 39$^{\circ}$ for the two cases of radial functions of outflow rate. The left and right panels correspond to the cases with $s=-0.5$, and $s= -1.5$, respectively. The line flux is normalized to one and expressed in arbitrary units. \label{fig:profile}}
\end{figure}

We use the velocity distributions of BLR clouds in FRADO model of mean quasar to calculate the broad emission line profiles based on the Doppler shift with respect to an observer's line of sight at mean viewing angle of 39$^{\circ}$ \citep{lawrence2010}.
We first consider a uniform emissivity for the BLR clouds, regardless of their location. The emission line profiles are thus computed by applying the Doppler shift to get the observed wavelength of each cloud. The results of line shape modeling are presented in figure \ref{fig:profile}.

\begin{figure}[H]
\begin{adjustwidth}{-\extralength}{0cm}
\centering
\includegraphics[width=16cm]{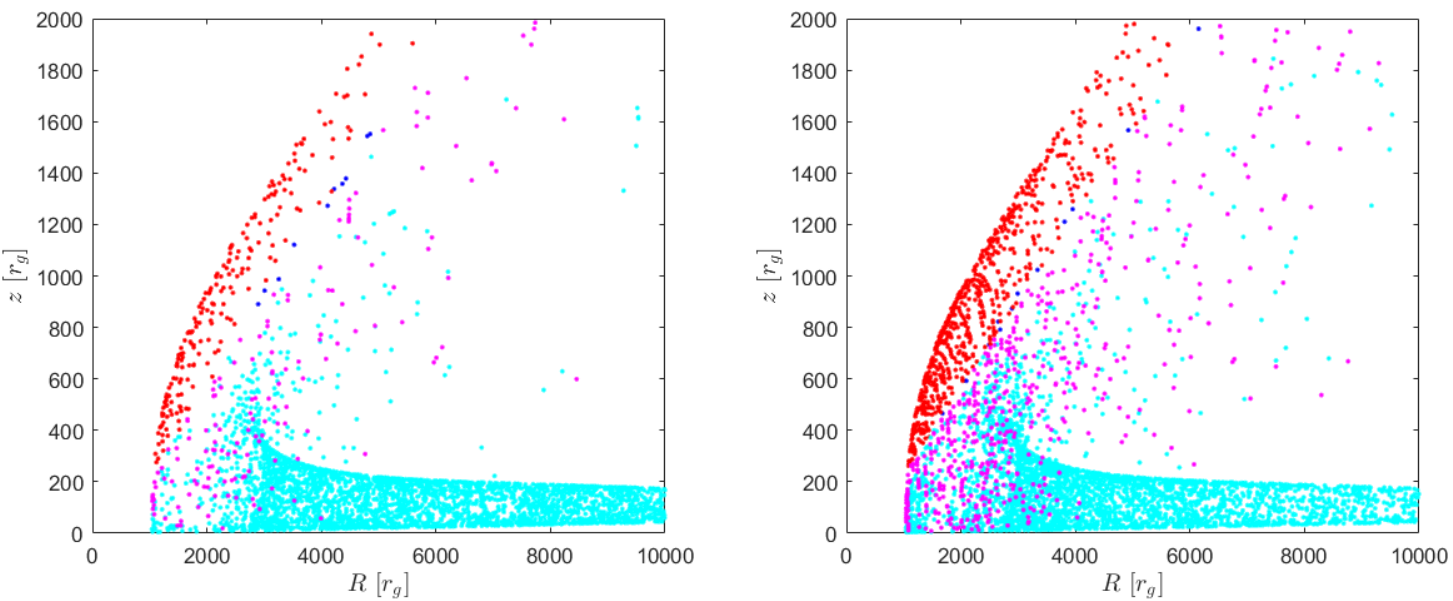}
\end{adjustwidth}
\caption{Spatial distribution of BLR clouds for the same model as in figure \ref{fig:distribution}, though distinguished with the photon flux. Dusty and dustless clouds are shown in blue and red, respectively (if their photon flux $\phi > \sim 10^{18}$ cm$^{-2}$ s$^{-1}$); otherwise shown in cyan and magenta, respectively. \label{fig:distribution1e17}}
\end{figure}

\begin{figure}[H]
\begin{adjustwidth}{-\extralength}{0cm}
\centering
\includegraphics[width=16cm]{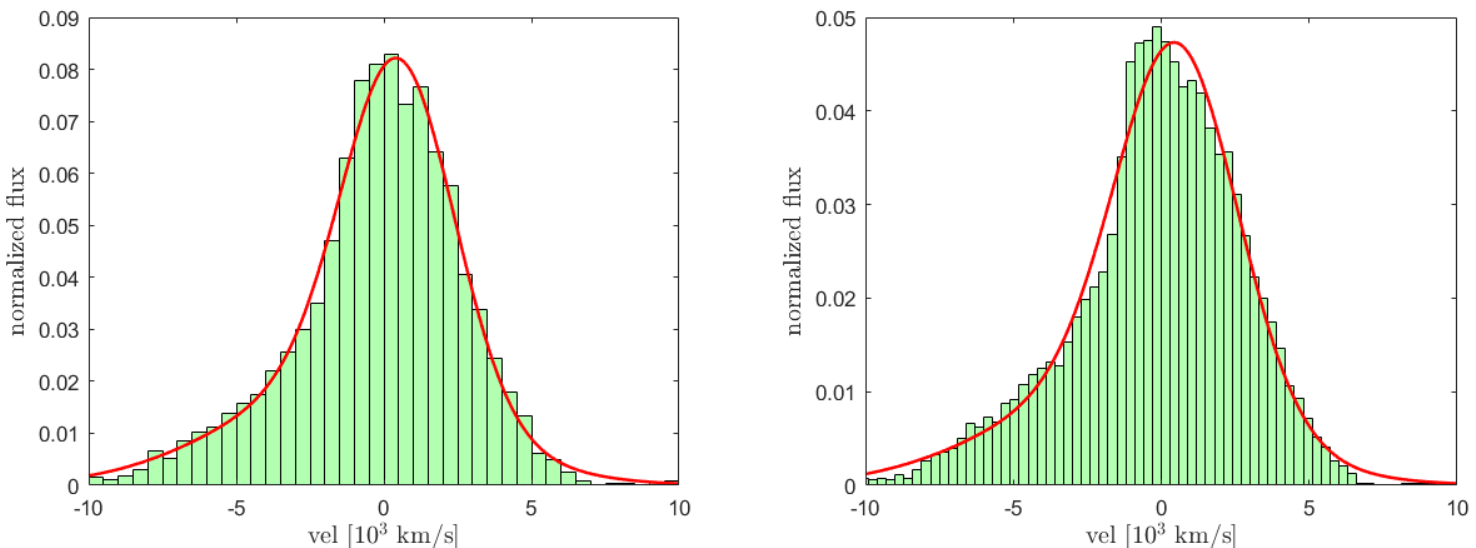}
\end{adjustwidth}
\caption{The emission line profiles for the same model as in figure \ref{fig:profile}; but with the new condition for emissivity of clouds based on the photon flux. \label{fig:profile_1e17}}
\end{figure}

We then introduce a physical condition which sets a threshold for clouds to contribute in the line shape. The condition comes from the parameter photon flux, $\phi$, which is also directly computed by the 2.5 FRADO. It depends on the location of a cloud and is required to be for example $\log(\phi) > \sim 17-18$ photons cm$^{-2}$ s$^{-1}$ to give rise to formation of H$\beta$ or MgII emission lines \citep{pandey2023}. Therefore, in the second step we still consider uniform emissivity but only include those clouds which meet the condition.
The clouds meeting the condition are distinguished in figure \ref{fig:distribution1e17}.
The line profiles with respect to the new condition are represented in figure \ref{fig:profile_1e17}.

\section{Discussion}\label{sec:discussion}

We considered two cases for the radial behavior of outflow from the disk, which contributes to the formation of the low-ionization BLR. Based on these two scenarios, we constructed the phase-space distribution of BLR clouds, representing their spatial and velocity distributions. Using this framework, we then calculated the shape of emission line profiles resulting from the Doppler shift of clouds under two distinct assumptions about their emissivity. In the first scenario, we assumed a uniform emissivity, where all clouds contribute equally to the line formation regardless of their local conditions. In the second scenario, we adopted a more physically motivated approach, allowing only those clouds receiving a photon flux of $\phi > \sim 10^{18}$ cm$^{-2}$ s$^{-1}$ to contribute to the emission. This threshold reflects the minimum radiation necessary to efficiently ionize and excite the clouds, thus influencing the line formation and overall profile shape. The results demonstrated the impact of different radial functions of disk outflow on both the geometry of the BLR and the observable emission line shapes.

As for the first scenario with single emissivity for all clouds, as in figure \ref{fig:profile}, the line profile with $s=-0.5$ exhibits a clear double-peaked structure with the peaks symmetrically distributed around the zero-velocity center and a dip which indicates reduced emission at low line-of-sight velocities, i.e. low vertical velocities. But the line profile with steeper outflow is more inclined to single-peaked and is comparatively smoother; the central dip observed in the left panel is filled in, as the emission from material at larger radii contributes less in number to the line profile, resulting in a more uniform emission profile around zero velocity. Double-peaked profiles are more characteristic of rotating disk-like systems, so the outflow rate with $s=-1.5$ aligns better with the observationally inferred dynamics of the BLR with respect to the mean quasar model, where gas closer to the central source dominates the emission. The steepest decline in the radial scaling of the disk outflow rate as $r^{-3/2}$ is then more favored in reproducing the overall shape of the low-ionization broad emission profile. This scaling behavior, intriguingly, also mirrors the radial dependency of the local Keplerian orbital time of BLR clouds in the gravitational potential of the central black hole, where $1/T \propto r^{-3/2}$, which in other words may reflect the frequency of clouds impacting the disk. If the clouds, after completing their orbital motion, return to the disk surface as part of a failed wind, their impact could perturb the disk surface. The turbulence generated at the impact site may, in turn, facilitate the launch of another cloud from the same location. While not all clouds are expected to return to the disk surface, this similarity in scaling laws is compelling, suggesting that the processes governing the outflow launching radial efficiency may be linked to the timescales of orbital motion. Numerical simulations \citep{proga2004, waters2016} support the idea that failed winds \citep{czerny2011} can significantly influence the disk environment.

When it comes to second scenario of emissivity where a more physically motivated approach was introduced by applying a photon flux threshold, both models of the radial scaling of the outflow produce a similar single-peaked line profile, as shown in figure \ref{fig:profile_1e17}. This result demonstrates that the new approach naturally suppresses contributions from clouds at larger radii with insufficient illumination, effectively simplifying the line shape. The primary reason for this similarity of line profiles here is not only that the line formation in both cases of outflow radial scaling is predominantly influenced by clouds at smaller radii, closer to the black hole, as shown figure \ref{fig:distribution1e17}, resulting in a single-peaked profile; but also that the clouds contributing to the line are launched from a very narrow radial range of the accretion disk (direct implication of new condition based on photon flux). This narrow launching range limits the sensitivity of the line shape to the specific radial scaling of the outflow. Consequently, while the steeper outflow function ($s=-1.5$) is preferred when considering the single emissivity model (as it better reproduces observed profiles, see figure \ref{fig:profile}), the photon-flux-limited approach diminishes the impact of the outflow radial function on the resulting line profile. This suggests that the emission line shapes in the flux-threshold scenario are largely dictated by the dynamics and illumination conditions of clouds in the innermost regions, regardless of the specific outflow rate scaling.

Upon transition to the inner region where the temperature exceeds that of dust sublimation, which is dominated by line driving mechanism, calculation of the total mass loss rate of the wind (fully escaping outflow) in \citep{murray1995, MurrayChiang1995} yields a relation as $\dot{M}_z \propto (L/L_{\rm Edd})^{1/\alpha}$ where $\alpha$ is reported to be $0.9$ for quasars, and $0.6$ for Seyferts.
This $\alpha$ relation indicates that the total outflow assuming the same luminosity is stronger in Seyferts than quasars, however quasars are in general much more luminous than Seyferts by orders of magnitudes.
The relation, replacing for $L(r) \propto r^{-2}$, can be rearranged to the power-law form of interest of this work as $\dot{M}_z(r) \propto r^{s}$ with $s = -1/2\alpha$ of around $-2.22$ and $-3.33$ for quasars and Seyferts, respectively. This means that the decline in the outflow launching efficiency of the disk in radial direction is much steeper for Seyferts than quasars, though not representing the net outflow per given radius as we do not take care of proportionality constant in the function $L(r)$. These $s$ indices below $-2$ in line-driving region, responsible for the formation of high-ionization broad emission lines, indicates that upon the transition from dusty region (where FRADO works) to inner dust-sublimating regions the outflow launching efficiency drops much steeper compared to the outer dust dominated disk so that the high-ionization outflow are more concentrated and massive at radii closer in to the black hole, i.e. at inner high-ionization BLR.

The velocities shown in the line profile figures are given with respect to the observer's reference frame. The asymmetry in the line shape, shifted toward negative velocities (blueward), indicates the presence of an escaping outflow, although this effect is not strongly pronounced. The full width at half maximum (FWHM) of the line profiles in all cases, as can be seen in figures \ref{fig:profile} and \ref{fig:profile_1e17}, is approximately $\sim 5000 - 6000$ km/s. With respect to definition of the gravitational radius, $r_{\rm g} = G M / c^2$, we can write $f R_{\rm BLR} = (c/{\rm FWHM})^2 r_{\rm g}$. Assuming a virial factor of $f=1$, this FWHM corresponds to a BLR size of $R_{\rm BLR} \sim 3000~r_{\rm g}$. This estimate aligns with the effective location of the emitting region shown in figure \ref{fig:distribution1e17}, suggesting that the BLR in this case is close to a virialized state.

In order to position the FRADO model of the mean quasar on the Eigenvector 1 (EV1) diagram, several important aspects must be considered. First, regarding the line emissivity and profile shape, the results from the first scenario (figure \ref{fig:profile}) exhibit double-peaked profiles with FWHM $> 4000$ km/s, which aligns with the behavior of Population B AGNs. However, the presence of blue asymmetry in these profiles is atypical for Pop B1 (or column 1 of the EV1 diagram in general, i.e. low Eddington ratio sources), as these populations more commonly exhibit redward asymmetry due to gravitational redshift effects \citep{sulentic2002, bon2015}. On the other hand, the second scenario (figure \ref{fig:profile_1e17}) shows single-peaked profiles with a slight blueshift, which is more typical of Pop A2-to-A4 objects. Despite this, the FWHM remains $> 4000$ km/s, which is characteristic of Pop B behavior. To evaluate the role of gravitational redshift, we follow \citep{ZhengSulentic1990}, where $z_{\rm grav} = r_{\rm g} / R_{\rm BLR} $. For the FRADO model of a mean quasar with $R_{\rm BLR} \sim 3000~r_{\rm g}$, the gravitational redshift is approximately 0.0003, a negligible value that does not significantly affect the line shape. This arises because the mean quasar has a moderate Eddington ratio of $\sim$ 0.1, placing its low-ionization BLR (e.g., H$\beta$) at a larger distance from the black hole compared to that of lower Eddington ratio sources, where relativistic effects are more pronounced.
Altogether, regardless of the chosen emissivity approach (resulting in single- or double-peaked line shapes), the combination of (a) FWHM $> 4000$ km/s, (b) the presence of blue-shifted asymmetry (a signature of outflow), and (c) the absence of redward boosted wings, suggests that the FRADO model of the mean quasar can be classified in the B2 tile of the EV1 diagram \citep{sulentic2008}. This classification aligns with a moderate accretion rate, a narrow stream of escaping outflow \citep{naddaf2022}, and negligible relativistic effects due to the distant location of the low-ionization BLR.
In future, a more sophisticated treatment of the cloud emissivity (for example, using detailed radiative transfer codes), and incorporating the effect of turbulence may significantly influence the final line profile shapes. Turbulence is fundamentally responsible for the Lorentzian bases observed in typical AGN line profiles \citep{kollatschny2013}. However, these early results suggest that even with an inner BLR highly populated with turbulent clouds to enhance the line's core, it is unlikely that the FWHM would decrease sufficiently to relocate the FRADO model of mean quasar into Pop A. Therefore, while refinements may improve alignment with observations, the general classification in Pop B2 remains robust for the mean quasar model.

\section{Conclusions}\label{sec:conclusion}

By comparing two distinct radial outflow rate scaling scenarios within the framework of the FRADO model, we demonstrated how the mass ejection rate of the disk influences the BLR geometry, and the observed line profiles. We showed that the steeper outflow scaling as $r^{-3/2}$ yields a more compact BLR with higher-velocity clouds concentrated near the black hole, producing a more centrally peaked profile. Although, with simplified examples we showed that the approach to the emissivity of BLR clouds turns to be very important in shaping the line profile, the $r^{-3/2}$ scaling is more favored. 
The outflow launching radial efficiency in high-ionization BLR follows a power-law with an index below $-2$ down to around $\sim -3.5$ and upon the transition to the dust-dominated region of the disk (FRADO) becomes shallower to an index of around $\sim -1.5$.
Moreover, there are many other physical parameters and phenomena which may affect the final shape of the emission lines which are overlooked in this work. Radiative transfer effects also play a crucial role in shaping the overall line profile. Therefore, simply knowing the emissivity profile of emission lines within the BLR is insufficient for accurately characterizing their shapes \citep{matthews202}. We will examine the properties of line profiles under different conditions and governing mechanisms in near future with a large grid of simulations with 2.5D FRADO model [Naddaf et al. 2025, in preparation] along with radiative transfer considerations [Savic \& Naddaf et al. 2025, in preparation].

\vspace{6pt}

\funding{This research was supported by the University of Liege under Special Funds for Research, IPD-STEMA Program.}

\begin{adjustwidth}{-\extralength}{0cm}
\reftitle{References}

\externalbibliography{yes}
\bibliography{naddaf.bib}

\begin{thebibliography}{999}

\bibitem[{Schmidt}(1963)]{schmidt1963}
{Schmidt}, M.
\newblock {3C 273 : A Star-Like Object with Large Red-Shift}.
\newblock {\em \nat} {\bf 1963}, {\em 197},~1040.
\newblock {\url{https://doi.org/10.1038/1971040a0}}.

\bibitem[{Antonucci}(1993)]{antonucci1993}
{Antonucci}, R.
\newblock {Unified models for active galactic nuclei and quasars}.
\newblock {\em ARA\&A} {\bf 1993}, {\em 31},~473--521.
\newblock {\url{https://doi.org/10.1146/annurev.aa.31.090193.002353}}.

\bibitem[{Netzer}(2015)]{netzer2015}
{Netzer}, H.
\newblock {Revisiting the Unified Model of Active Galactic Nuclei}.
\newblock {\em ARA\&A} {\bf 2015}, {\em 53},~365--408,  \href{http://xxx.lanl.gov/abs/1505.00811}{{\normalfont [1505.00811]}}.
\newblock {\url{https://doi.org/10.1146/annurev-astro-082214-122302}}.

\bibitem[{Ramos Almeida} and {Ricci}(2017)]{Almeida2017}
{Ramos Almeida}, C.; {Ricci}, C.
\newblock {Nuclear obscuration in active galactic nuclei}.
\newblock {\em Nature Astronomy} {\bf 2017}, {\em 1},~679--689,  \href{http://xxx.lanl.gov/abs/1709.00019}{{\normalfont [arXiv:astro-ph.GA/1709.00019]}}.
\newblock {\url{https://doi.org/10.1038/s41550-017-0232-z}}.

\bibitem[{Gaskell}(2009)]{gaskell2009}
{Gaskell}, C.M.
\newblock {What broad emission lines tell us about how active galactic nuclei work}.
\newblock {\em \nar} {\bf 2009}, {\em 53},~140--148,  \href{http://xxx.lanl.gov/abs/0908.0386}{{\normalfont [0908.0386]}}.
\newblock {\url{https://doi.org/10.1016/j.newar.2009.09.006}}.

\bibitem[{Collin-Souffrin} et~al.(1988){Collin-Souffrin}, {Dyson}, {McDowell}, and {Perry}]{CollinSouffrin1988}
{Collin-Souffrin}, S.; {Dyson}, J.E.; {McDowell}, J.C.; {Perry}, J.J.
\newblock {The environment of active galactic nuclei. I - A two-component broad emission line model}.
\newblock {\em MNRAS} {\bf 1988}, {\em 232},~539--550.
\newblock {\url{https://doi.org/10.1093/mnras/232.3.539}}.

\bibitem[{Kollatschny}(2003)]{kollatschny2003}
{Kollatschny}, W.
\newblock {Accretion disk wind in the AGN broad-line region: Spectroscopically resolved line profile variations in Mrk 110}.
\newblock {\em \aap} {\bf 2003}, {\em 407},~461--472,  \href{http://xxx.lanl.gov/abs/astro-ph/0306389}{{\normalfont [arXiv:astro-ph/astro-ph/0306389]}}.
\newblock {\url{https://doi.org/10.1051/0004-6361:20030928}}.

\bibitem[{Czerny} et~al.(2015){Czerny}, {Modzelewska}, {Petrogalli}, {Pych}, {Adhikari}, {{\.Z}ycki}, {Hryniewicz}, {Krupa}, {{\'S}wie{\c{t}}o{\'n}}, and {Niko{\l}ajuk}]{czerny2015}
{Czerny}, B.; {Modzelewska}, J.; {Petrogalli}, F.; {Pych}, W.; {Adhikari}, T.P.; {{\.Z}ycki}, P.T.; {Hryniewicz}, K.; {Krupa}, M.; {{\'S}wie{\c{t}}o{\'n}}, A.; {Niko{\l}ajuk}, M.
\newblock {The dust origin of the Broad Line Region and the model consequences for AGN unification scheme}.
\newblock {\em Advances in Space Research} {\bf 2015}, {\em 55},~1806--1815,  \href{http://xxx.lanl.gov/abs/1409.7312}{{\normalfont [arXiv:astro-ph.GA/1409.7312]}}.
\newblock {\url{https://doi.org/10.1016/j.asr.2015.01.004}}.

\bibitem[{Leighly} and {Casebeer}(2007)]{leighly2007}
{Leighly}, K.M.; {Casebeer}, D.
\newblock {Photoionization Models of the Broad-line Region}.
\newblock In Proceedings of the The Central Engine of Active Galactic Nuclei; {Ho}, L.C.; {Wang}, J.W., Eds.,  2007, Vol. 373, {\em Astronomical Society of the Pacific Conference Series}, p. 365,  \href{http://xxx.lanl.gov/abs/astro-ph/0701437}{{\normalfont [arXiv:astro-ph/astro-ph/0701437]}}.
\newblock {\url{https://doi.org/10.48550/arXiv.astro-ph/0701437}}.

\bibitem[{Collin} et~al.(2006){Collin}, {Kawaguchi}, {Peterson}, and {Vestergaard}]{collin2006}
{Collin}, S.; {Kawaguchi}, T.; {Peterson}, B.M.; {Vestergaard}, M.
\newblock {Systematic effects in measurement of black hole masses by emission-line reverberation of active galactic nuclei: Eddington ratio and inclination}.
\newblock {\em \aap} {\bf 2006}, {\em 456},~75--90,  \href{http://xxx.lanl.gov/abs/astro-ph/0603460}{{\normalfont [arXiv:astro-ph/astro-ph/0603460]}}.
\newblock {\url{https://doi.org/10.1051/0004-6361:20064878}}.

\bibitem[{Gravity Collaboration} et~al.(2018){Gravity Collaboration}, {Sturm}, {Dexter}, {Pfuhl}, {Stock}, {Davies}, {Lutz}, {Cl{\'e}net}, {Eckart}, {Eisenhauer}, {Genzel}, {Gratadour}, {H{\"o}nig}, {Kishimoto}, {Lacour}, {Millour}, {Netzer}, {Perrin}, {Peterson}, {Petrucci}, {Rouan}, {Waisberg}, {Woillez}, {Amorim}, {Brandner}, {F{\"o}rster Schreiber}, {Garcia}, {Gillessen}, {Ott}, {Paumard}, {Perraut}, {Scheithauer}, {Straubmeier}, {Tacconi}, and {Widmann}]{GravityColl2018}
{Gravity Collaboration}.; {Sturm}, E.; {Dexter}, J.; {Pfuhl}, O.; {Stock}, M.R.; {Davies}, R.I.; {Lutz}, D.; {Cl{\'e}net}, Y.; {Eckart}, A.; {Eisenhauer}, F.;  et~al.
\newblock {Spatially resolved rotation of the broad-line region of a quasar at sub-parsec scale}.
\newblock {\em \nat} {\bf 2018}, {\em 563},~657--660,  \href{http://xxx.lanl.gov/abs/1811.11195}{{\normalfont [arXiv:astro-ph.GA/1811.11195]}}.
\newblock {\url{https://doi.org/10.1038/s41586-018-0731-9}}.

\bibitem[{H{\"o}nig} et~al.(2013){H{\"o}nig}, {Kishimoto}, {Tristram}, {Prieto}, {Gandhi}, {Asmus}, {Antonucci}, {Burtscher}, {Duschl}, and {Weigelt}]{honig2013}
{H{\"o}nig}, S.F.; {Kishimoto}, M.; {Tristram}, K.R.W.; {Prieto}, M.A.; {Gandhi}, P.; {Asmus}, D.; {Antonucci}, R.; {Burtscher}, L.; {Duschl}, W.J.; {Weigelt}, G.
\newblock {Dust in the Polar Region as a Major Contributor to the Infrared Emission of Active Galactic Nuclei}.
\newblock {\em \apj} {\bf 2013}, {\em 771},~87,  \href{http://xxx.lanl.gov/abs/1306.4312}{{\normalfont [arXiv:astro-ph.CO/1306.4312]}}.
\newblock {\url{https://doi.org/10.1088/0004-637X/771/2/87}}.

\bibitem[{Elitzur} and {Shlosman}(2006)]{elitzur2006}
{Elitzur}, M.; {Shlosman}, I.
\newblock {The AGN-obscuring Torus: The End of the ``Doughnut'' Paradigm?}
\newblock {\em \apjl} {\bf 2006}, {\em 648},~L101--L104,  \href{http://xxx.lanl.gov/abs/astro-ph/0605686}{{\normalfont [arXiv:astro-ph/astro-ph/0605686]}}.
\newblock {\url{https://doi.org/10.1086/508158}}.

\bibitem[{Proga} et~al.(2000){Proga}, {Stone}, and {Kallman}]{proga2000}
{Proga}, D.; {Stone}, J.M.; {Kallman}, T.R.
\newblock {Dynamics of Line-driven Disk Winds in Active Galactic Nuclei}.
\newblock {\em \apj} {\bf 2000}, {\em 543},~686--696,  \href{http://xxx.lanl.gov/abs/astro-ph/0005315}{{\normalfont [arXiv:astro-ph/astro-ph/0005315]}}.
\newblock {\url{https://doi.org/10.1086/317154}}.

\bibitem[{Naddaf} et~al.(2021){Naddaf}, {Czerny}, and {Szczerba}]{naddaf2021}
{Naddaf}, M.H.; {Czerny}, B.; {Szczerba}, R.
\newblock {The Picture of BLR in 2.5D FRADO: Dynamics and Geometry}.
\newblock {\em \apj} {\bf 2021}, {\em 920},~30,  \href{http://xxx.lanl.gov/abs/2102.00336}{{\normalfont [arXiv:astro-ph.GA/2102.00336]}}.
\newblock {\url{https://doi.org/10.3847/1538-4357/ac139d}}.

\bibitem[{Netzer} and {Laor}(1993)]{netzer1993}
{Netzer}, H.; {Laor}, A.
\newblock {Dust in the Narrow-Line Region of Active Galactic Nuclei}.
\newblock {\em \apjl} {\bf 1993}, {\em 404},~L51.
\newblock {\url{https://doi.org/10.1086/186741}}.

\bibitem[{Nicastro}(2000)]{nicastro2000}
{Nicastro}, F.
\newblock {Broad Emission Line Regions in Active Galactic Nuclei: The Link with the Accretion Power}.
\newblock {\em \apjl} {\bf 2000}, {\em 530},~L65--L68,  \href{http://xxx.lanl.gov/abs/astro-ph/9912524}{{\normalfont [arXiv:astro-ph/astro-ph/9912524]}}.
\newblock {\url{https://doi.org/10.1086/312491}}.

\bibitem[{Czerny} and {Hryniewicz}(2011)]{czerny2011}
{Czerny}, B.; {Hryniewicz}, K.
\newblock {The origin of the broad line region in active galactic nuclei}.
\newblock {\em \aap} {\bf 2011}, {\em 525},~L8,  \href{http://xxx.lanl.gov/abs/1010.6201}{{\normalfont [arXiv:astro-ph.CO/1010.6201]}}.
\newblock {\url{https://doi.org/10.1051/0004-6361/201016025}}.

\bibitem[{Naddaf} and {Czerny}(2021)]{naddaf2022}
{Naddaf}, M.H.; {Czerny}, B.
\newblock {Radiation pressure on dust explains the Low Ionized Broad Emission Lines in Active Galactic Nuclei}.
\newblock {\em arXiv e-prints} {\bf 2021}, p. arXiv:2111.14963,  \href{http://xxx.lanl.gov/abs/2111.14963}{{\normalfont [arXiv:astro-ph.GA/2111.14963]}}.

\bibitem[{Murray} et~al.(1995){Murray}, {Chiang}, {Grossman}, and {Voit}]{murray1995}
{Murray}, N.; {Chiang}, J.; {Grossman}, S.A.; {Voit}, G.M.
\newblock {Accretion Disk Winds from Active Galactic Nuclei}.
\newblock {\em \apj} {\bf 1995}, {\em 451},~498.
\newblock {\url{https://doi.org/10.1086/176238}}.

\bibitem[{Czerny} et~al.(2016){Czerny}, {Du}, {Wang}, and {Karas}]{czerny2016}
{Czerny}, B.; {Du}, P.; {Wang}, J.M.; {Karas}, V.
\newblock {A Test of the Formation Mechanism of the Broad Line Region in Active Galactic Nuclei}.
\newblock {\em ApJ} {\bf 2016}, {\em 832},~15,  \href{http://xxx.lanl.gov/abs/1610.00420}{{\normalfont [arXiv:astro-ph.GA/1610.00420]}}.
\newblock {\url{https://doi.org/10.3847/0004-637X/832/1/15}}.

\bibitem[{Czerny} et~al.(2017){Czerny}, {Li}, {Hryniewicz}, {Panda}, {Wildy}, {Sniegowska}, {Wang}, {Sredzinska}, and {Karas}]{czerny2017}
{Czerny}, B.; {Li}, Y.R.; {Hryniewicz}, K.; {Panda}, S.; {Wildy}, C.; {Sniegowska}, M.; {Wang}, J.M.; {Sredzinska}, J.; {Karas}, V.
\newblock {Failed Radiatively Accelerated Dusty Outflow Model of the Broad Line Region in Active Galactic Nuclei. I. Analytical Solution}.
\newblock {\em ApJ} {\bf 2017}, {\em 846},~154,  \href{http://xxx.lanl.gov/abs/1706.07958}{{\normalfont [arXiv:astro-ph.GA/1706.07958]}}.
\newblock {\url{https://doi.org/10.3847/1538-4357/aa8810}}.

\bibitem[{Rees} et~al.(1969){Rees}, {Silk}, {Werner}, and {Wickramasinghe}]{rees1969}
{Rees}, M.J.; {Silk}, J.I.; {Werner}, M.W.; {Wickramasinghe}, N.C.
\newblock {Infrared Radiation from Dust in Seyfert Galaxies}.
\newblock {\em \nat} {\bf 1969}, {\em 223},~788--791.
\newblock {\url{https://doi.org/10.1038/223788a0}}.

\bibitem[{Dong} et~al.(2008){Dong}, {Wang}, {Wang}, {Yuan}, {Zhou}, {Dai}, and {Zhang}]{dong2008}
{Dong}, X.; {Wang}, T.; {Wang}, J.; {Yuan}, W.; {Zhou}, H.; {Dai}, H.; {Zhang}, K.
\newblock {Broad-line Balmer decrements in blue active galactic nuclei}.
\newblock {\em MNRAS} {\bf 2008}, {\em 383},~581--592,  \href{http://xxx.lanl.gov/abs/0710.1458}{{\normalfont [arXiv:astro-ph/0710.1458]}}.
\newblock {\url{https://doi.org/10.1111/j.1365-2966.2007.12560.x}}.

\bibitem[{Naddaf} et~al.(2020){Naddaf}, {Czerny}, and {Szczerba}]{naddaf2020}
{Naddaf}, M.H.; {Czerny}, B.; {Szczerba}, R.
\newblock {BLR size in Realistic FRADO Model}.
\newblock {\em Frontiers in Astronomy and Space Sciences} {\bf 2020}, {\em 7},~15,  \href{http://xxx.lanl.gov/abs/1912.00278}{{\normalfont [arXiv:astro-ph.HE/1912.00278]}}.
\newblock {\url{https://doi.org/10.3389/fspas.2020.00015}}.

\bibitem[{Shakura} and {Sunyaev}(1973)]{SS73}
{Shakura}, N.I.; {Sunyaev}, R.A.
\newblock {Black holes in binary systems. Observational appearance.}
\newblock {\em A\&A} {\bf 1973}, {\em 500},~33--51.

\bibitem[{Naddaf} et~al.(2022){Naddaf}, {Czerny}, and {Zaja{\v{c}}ek}]{naddaf2022Dynam}
{Naddaf}, M.H.; {Czerny}, B.; {Zaja{\v{c}}ek}, M.
\newblock {The Wind Dynamics of Super-Eddington Sources in FRADO}.
\newblock {\em Dynamics} {\bf 2022}, {\em 2},~295--305,  \href{http://xxx.lanl.gov/abs/2209.09304}{{\normalfont [arXiv:astro-ph.GA/2209.09304]}}.
\newblock {\url{https://doi.org/10.3390/dynamics2030015}}.

\bibitem[{Murray} and {Chiang}(1995)]{MurrayChiang1995}
{Murray}, N.; {Chiang}, J.
\newblock {Active Galactic Nuclei Disk Winds, Absorption Lines, and Warm Absorbers}.
\newblock {\em \apjl} {\bf 1995}, {\em 454},~L105.
\newblock {\url{https://doi.org/10.1086/309775}}.

\bibitem[{Castor} et~al.(1975){Castor}, {Abbott}, and {Klein}]{castor1975}
{Castor}, J.I.; {Abbott}, D.C.; {Klein}, R.I.
\newblock {Radiation-driven winds in Of stars.}
\newblock {\em \apj} {\bf 1975}, {\em 195},~157--174.
\newblock {\url{https://doi.org/10.1086/153315}}.

\bibitem[{Muijres} et~al.(2011){Muijres}, {de Koter}, {Vink}, {Krti{\v{c}}ka}, {Kub{\'a}t}, and {Langer}]{muijres2011}
{Muijres}, L.E.; {de Koter}, A.; {Vink}, J.S.; {Krti{\v{c}}ka}, J.; {Kub{\'a}t}, J.; {Langer}, N.
\newblock {Predictions of the effect of clumping on the wind properties of O-type stars}.
\newblock {\em \aap} {\bf 2011}, {\em 526},~A32.
\newblock {\url{https://doi.org/10.1051/0004-6361/201014290}}.

\bibitem[{Arav} and {Li}(1994)]{arav1994}
{Arav}, N.; {Li}, Z.Y.
\newblock {The Role of Radiative Acceleration in Outflows from Broad Absorption Line QSOs. I. Comparison with O Star Winds}.
\newblock {\em \apj} {\bf 1994}, {\em 427},~700.
\newblock {\url{https://doi.org/10.1086/174177}}.

\bibitem[{Lamers} et~al.(1995){Lamers}, {Snow}, and {Lindholm}]{lamers1995}
{Lamers}, H.J.G.L.M.; {Snow}, T.P.; {Lindholm}, D.M.
\newblock {Terminal Velocities and the Bistability of Stellar Winds}.
\newblock {\em \apj} {\bf 1995}, {\em 455},~269.
\newblock {\url{https://doi.org/10.1086/176575}}.

\bibitem[{Semenov} et~al.(2003){Semenov}, {Henning}, {Helling}, {Ilgner}, and {Sedlmayr}]{Semenov2003}
{Semenov}, D.; {Henning}, T.; {Helling}, C.; {Ilgner}, M.; {Sedlmayr}, E.
\newblock {Rosseland and Planck mean opacities for protoplanetary discs}.
\newblock {\em \aap} {\bf 2003}, {\em 410},~611--621,  \href{http://xxx.lanl.gov/abs/astro-ph/0308344}{{\normalfont [arXiv:astro-ph/astro-ph/0308344]}}.
\newblock {\url{https://doi.org/10.1051/0004-6361:20031279}}.

\bibitem[{Vanden Berk} et~al.(2001){Vanden Berk}, {Richards}, {Bauer}, {Strauss}, {Schneider}, {Heckman}, {York}, {Hall}, {Fan}, {Knapp}, {Anderson}, {Annis}, {Bahcall}, {Bernardi}, {Briggs}, {Brinkmann}, {Brunner}, {Burles}, {Carey}, {Castander}, {Connolly}, {Crocker}, {Csabai}, {Doi}, {Finkbeiner}, {Friedman}, {Frieman}, {Fukugita}, {Gunn}, {Hennessy}, {Ivezi{\'c}}, {Kent}, {Kunszt}, {Lamb}, {Leger}, {Long}, {Loveday}, {Lupton}, {Meiksin}, {Merelli}, {Munn}, {Newberg}, {Newcomb}, {Nichol}, {Owen}, {Pier}, {Pope}, {Rockosi}, {Schlegel}, {Siegmund}, {Smee}, {Snir}, {Stoughton}, {Stubbs}, {SubbaRao}, {Szalay}, {Szokoly}, {Tremonti}, {Uomoto}, {Waddell}, {Yanny}, and {Zheng}]{vandenberk2001}
{Vanden Berk}, D.E.; {Richards}, G.T.; {Bauer}, A.; {Strauss}, M.A.; {Schneider}, D.P.; {Heckman}, T.M.; {York}, D.G.; {Hall}, P.B.; {Fan}, X.; {Knapp}, G.R.;  et~al.
\newblock {Composite Quasar Spectra from the Sloan Digital Sky Survey}.
\newblock {\em \aj} {\bf 2001}, {\em 122},~549--564,  \href{http://xxx.lanl.gov/abs/astro-ph/0105231}{{\normalfont [arXiv:astro-ph/astro-ph/0105231]}}.
\newblock {\url{https://doi.org/10.1086/321167}}.

\bibitem[{Shen} et~al.(2011){Shen}, {Richards}, {Strauss}, {Hall}, {Schneider}, {Snedden}, {Bizyaev}, {Brewington}, {Malanushenko}, {Malanushenko}, {Oravetz}, {Pan}, and {Simmons}]{shen2011}
{Shen}, Y.; {Richards}, G.T.; {Strauss}, M.A.; {Hall}, P.B.; {Schneider}, D.P.; {Snedden}, S.; {Bizyaev}, D.; {Brewington}, H.; {Malanushenko}, V.; {Malanushenko}, E.;  et~al.
\newblock {A Catalog of Quasar Properties from Sloan Digital Sky Survey Data Release 7}.
\newblock {\em ApJs} {\bf 2011}, {\em 194},~45,  \href{http://xxx.lanl.gov/abs/1006.5178}{{\normalfont [arXiv:astro-ph.CO/1006.5178]}}.
\newblock {\url{https://doi.org/10.1088/0067-0049/194/2/45}}.

\bibitem[{Lawrence} and {Elvis}(2010)]{lawrence2010}
{Lawrence}, A.; {Elvis}, M.
\newblock {Misaligned Disks as Obscurers in Active Galaxies}.
\newblock {\em \apj} {\bf 2010}, {\em 714},~561--570.
\newblock {\url{https://doi.org/10.1088/0004-637X/714/1/561}}.

\bibitem[{Pandey} et~al.(2023){Pandey}, {Czerny}, {Panda}, {Prince}, {Jaiswal}, {Martinez-Aldama}, {Zaja{\v{c}}ek}, and {{\'S}niegowska}]{pandey2023}
{Pandey}, A.; {Czerny}, B.; {Panda}, S.; {Prince}, R.; {Jaiswal}, V.K.; {Martinez-Aldama}, M.L.; {Zaja{\v{c}}ek}, M.; {{\'S}niegowska}, M.
\newblock {Broad-line region in active galactic nuclei: Dusty or dustless?}
\newblock {\em \aap} {\bf 2023}, {\em 680},~A102,  \href{http://xxx.lanl.gov/abs/2310.05089}{{\normalfont [arXiv:astro-ph.GA/2310.05089]}}.
\newblock {\url{https://doi.org/10.1051/0004-6361/202347819}}.

\bibitem[{Proga} and {Kallman}(2004)]{proga2004}
{Proga}, D.; {Kallman}, T.R.
\newblock {Dynamics of Line-driven Disk Winds in Active Galactic Nuclei. II. Effects of Disk Radiation}.
\newblock {\em \apj} {\bf 2004}, {\em 616},~688--695,  \href{http://xxx.lanl.gov/abs/astro-ph/0408293}{{\normalfont [arXiv:astro-ph/astro-ph/0408293]}}.
\newblock {\url{https://doi.org/10.1086/425117}}.

\bibitem[{Waters} et~al.(2016){Waters}, {Kashi}, {Proga}, {Eracleous}, {Barth}, and {Greene}]{waters2016}
{Waters}, T.; {Kashi}, A.; {Proga}, D.; {Eracleous}, M.; {Barth}, A.J.; {Greene}, J.
\newblock {Reverberation Mapping of the Broad Line Region: Application to a Hydrodynamical Line-driven Disk Wind Solution}.
\newblock {\em \apj} {\bf 2016}, {\em 827},~53,  \href{http://xxx.lanl.gov/abs/1601.05181}{{\normalfont [arXiv:astro-ph.GA/1601.05181]}}.
\newblock {\url{https://doi.org/10.3847/0004-637X/827/1/53}}.

\bibitem[{Sulentic} et~al.(2002){Sulentic}, {Marziani}, {Zamanov}, {Bachev}, {Calvani}, and {Dultzin-Hacyan}]{sulentic2002}
{Sulentic}, J.W.; {Marziani}, P.; {Zamanov}, R.; {Bachev}, R.; {Calvani}, M.; {Dultzin-Hacyan}, D.
\newblock {Average Quasar Spectra in the Context of Eigenvector 1}.
\newblock {\em \apjl} {\bf 2002}, {\em 566},~L71--L75,  \href{http://xxx.lanl.gov/abs/astro-ph/0201362}{{\normalfont [arXiv:astro-ph/astro-ph/0201362]}}.
\newblock {\url{https://doi.org/10.1086/339594}}.

\bibitem[{Bon} et~al.(2015){Bon}, {Bon}, {Marziani}, and {Jovanovi{\'c}}]{bon2015}
{Bon}, N.; {Bon}, E.; {Marziani}, P.; {Jovanovi{\'c}}, P.
\newblock {Gravitational redshift of emission lines in the AGN spectra}.
\newblock {\em \apss} {\bf 2015}, {\em 360},~7,  \href{http://xxx.lanl.gov/abs/1602.03688}{{\normalfont [arXiv:astro-ph.GA/1602.03688]}}.
\newblock {\url{https://doi.org/10.1007/s10509-015-2555-5}}.

\bibitem[{Zheng} and {Sulentic}(1990)]{ZhengSulentic1990}
{Zheng}, W.; {Sulentic}, J.W.
\newblock {Internal Redshift Difference and Central Mass in QSOs}.
\newblock {\em \apj} {\bf 1990}, {\em 350},~512.
\newblock {\url{https://doi.org/10.1086/168407}}.

\bibitem[{Sulentic} et~al.(2008){Sulentic}, {Zamfir}, {Marziani}, and {Dultzin}]{sulentic2008}
{Sulentic}, J.W.; {Zamfir}, S.; {Marziani}, P.; {Dultzin}, D.
\newblock {Our Search for an H-R Diagram of Quasars}.
\newblock In Proceedings of the Revista Mexicana de Astronomia y Astrofisica Conference Series,  2008, Vol.~32, {\em Revista Mexicana de Astronomia y Astrofisica Conference Series}, pp. 51--58,  \href{http://xxx.lanl.gov/abs/0709.2499}{{\normalfont [arXiv:astro-ph/0709.2499]}}.
\newblock {\url{https://doi.org/10.48550/arXiv.0709.2499}}.

\bibitem[{Kollatschny} and {Zetzl}(2013)]{kollatschny2013}
{Kollatschny}, W.; {Zetzl}, M.
\newblock {The shape of broad-line profiles in active galactic nuclei}.
\newblock {\em \aap} {\bf 2013}, {\em 549},~A100,  \href{http://xxx.lanl.gov/abs/1211.3065}{{\normalfont [arXiv:astro-ph.CO/1211.3065]}}.
\newblock {\url{https://doi.org/10.1051/0004-6361/201219411}}.

\bibitem[{Matthews} et~al.(2023){Matthews}, {Strong-Wright}, {Knigge}, {Hewett}, {Temple}, {Long}, {Rankine}, {Stepney}, {Banerji}, and {Richards}]{matthews202}
{Matthews}, J.H.; {Strong-Wright}, J.; {Knigge}, C.; {Hewett}, P.; {Temple}, M.J.; {Long}, K.S.; {Rankine}, A.L.; {Stepney}, M.; {Banerji}, M.; {Richards}, G.T.
\newblock {A disc wind model for blueshifts in quasar broad emission lines}.
\newblock {\em \mnras} {\bf 2023}, {\em 526},~3967--3986,  \href{http://xxx.lanl.gov/abs/2309.14434}{{\normalfont [arXiv:astro-ph.GA/2309.14434]}}.
\newblock {\url{https://doi.org/10.1093/mnras/stad2895}}.

\end{thebibliography}


\PublishersNote{}
\end{adjustwidth}
\end{document}